\documentclass[a4paper,11pt,hyper]{JHEP3}
\usepackage{amsfonts,latexsym,graphicx,epsfig,amssymb,amsmath,mathrsfs}

\newcommand{\bm}[1]{\hbox{\boldmath{$#1$}}}
\newcommand{\sbm}[1]{\hbox{\boldmath{\scriptsize$#1$}}}
\newcommand{\dd}{{\rm d}}
\newcommand{\Mp}{M_{\rm pl}}
\newcommand{\WQ}{W_{\rm QFT}}

\newcommand{\WI}{W^{(2)\,-1}}
\newcommand{\cp}{{\rm cyclic~perms}}

\newcommand{\cR}{{\cal R}}

\title{Holographic inflation and the conservation of $\zeta$}
\author{Jaume Garriga$^a$, Yuko Urakawa$^{a,b}$
\\
a. Departament de F{\'\i}sica Fonamental i Institut de Ci{\`e}ncies del Cosmos, 
Universitat de Barcelona,
Mart{\'\i}\ i Franqu{\`e}s 1, 08028 Barcelona, Spain\\
b. Department of Physics and Astrophysics, Nagoya University, Chikusa,
Nagoya 464-8602, Japan}

\abstract{In a holographic description of inflation, cosmological time
evolution in the bulk is expected to correspond to the renomalization group (RG)
flow in a dual boundary theory. Here, we analyze this expectation by
computing the correlation functions of the curvature perturbation $\zeta$ holographically.
For this purpose, we use a deformed conformal field theory at the
boundary, with a single deformation operator. In standard single field
models of inflation, $\zeta$ is known to be conserved at large scales
under very general conditions. However, we find that this is not
generically the case in the dual description. The requirement that
higher correlators of $\zeta$ should be conserved severely restricts the
possibilities for the RG flow. With such restriction, the power spectrum
$P_\zeta$ must follow an exact power law, at least within the regime of
validity of conformal perturbation theory.}

\keywords{Inflation, dS/CFT correspondence, Primordial perturbation}
\preprint{}

\maketitle

\begin{document}


\section{Introduction}   \label{Sec:Introduction}

Inflation has become a standard paradigm for describing the origin of
cosmological perturbations~\cite{Ade:2013rta, Ade:2014xna}. In fact, current observational data is in good agreement
with single field models, with just one inflaton field~\cite{Ade:2013rta}. On the other hand, it has been suggested that inflation may be described holographically by means of a dual field theory at the future boundary.
 According to the gauge/gravity correspondence, the strongly (weakly) coupled phase
of bulk gravity corresponds to the weakly (strongly) coupled phase of the dual boundary theory. 
Because of that, holography may open up new insights on the study of the very early universe, 
near the Planck scale, where non-perturbative gravitational effects may play  a role.

The gauge/gravity duality was initially advocated for asymptotically anti-de Sitter
(AdS) space times~\cite{Maldacena1997, GKP, Witten1998}. Such duality
cannot be immediately applied to inflationary cosmology, where the
spacetime is similar to de Sitter (dS) rather than AdS. Nonetheless, by
analogy, there have been several suggestions that a $(d+1)$-dimensional
inflationary evolution may be dual to a quantum field theory (QFT) on a
$d$-dimensional space with Euclidean signature. Following the work by
Strominger~\cite{Strominger, Strominger2} and Witten~\cite{Witten}, this
possibility has been further investigated in the context of dS~\cite{Bousso:2001mw, Harlow:2011ke, Anninos:2011ui} and
quasi-dS spacetimes~\cite{Strominger2, Maldacena02}. In
Refs.~\cite{Maldacena02, Seery:2006tq, vdS, LM03, LM04, Shiu,
Mata:2012bx, Ghosh:2014kba, JYsingle, Larsen:2014wpa}, the duality was discussed by
including cosmological perturbations. (See also
Refs.~\cite{Banks:2013qra, Banks:2013qpa} and Ref.~\cite{Pimentel:2013gza}.)
The holographic description of inflation has also been studied by using the so-called domain
wall/cosmology correspondence, where cosmological solutions are
constructed by analytically continuing from domain wall solutions~
\cite{MS_HC09, MS_HC10, MS_HCob10, MS_NG, MS_NGGW, BMS} (see also Ref.~\cite{Kiritsis:2013gia}).

The implementation of the duality requires a concrete dictionary, relating cosmological observables in the bulk
with field theory observables at the boundary. However, this relation
hasn't been clearly understood, except perhaps in certain limits, such
as the vicinity of a dS fixed point. In particular, it is not clear
which cosmological variable corresponds to the renormalization scale $\mu$. 
In Refs.~\cite{BMS, AJ08, AJ09, Alex11} , it was
argued that for the case of dS spacetime, $\mu$ should be proportional to the scale factor, $\mu \propto a$, but the
relations suggested in these references differ from each other when the solutions deviate from dS spacetime.

One may expect that in quasi-dS spacetimes the cosmological evolution in the bulk will be still
described by the renormalization group (RG) flow in the boundary. The purpose of this paper is to 
examine this naive expectation, by computing the evolution of the primordial
curvature perturbation $\zeta$. This plays an important role because in
standard cosmological perturbation theory (CPT) $\zeta$ is generically conserved for adiabatic perturbations at large
scales (this will be reviewed more precisely in Section 5). If the time
evolution in the bulk consistently corresponds to the RG flow in 
the dual boundary QFT, the correlators of $\zeta$ predicted in the boundary
QFT should be independent of $\mu$ in the large scale limit. In this
paper, we examine whether the RG flow in the boundary
QFT predicts that the correlators of $\zeta$ become $\mu$ independent or not.

The outline of this paper is as follows. In
Sec.~\ref{Sec:Preliminaries}, after we describe our setup, following
Ref.~\cite{JYsingle}, we provide a way to calculate the correlators of
$\zeta$ by using the wave function of the bulk spacetime. To consider
the boundary QFT which is dual to the single field inflation, we
introduce a single deformation term to the boundary
action, which lets the QFT deviate from the conformal field theory
(CFT). In Sec.~\ref{Sec:dCFT}, we discuss the solution of the RG
equation in the boundary theory. In Sec.~\ref{Sec:vertexfn}, using the
gauge transformation, we derive the relation between the correlators of
$\zeta$ and the correlators of the boundary operator. Then, in
Sec.~\ref{Sec:conservation}, computing the correlators of 
$\zeta$, we investigate their $\mu$ dependence at large scales.

\section{Preliminaries}  \label{Sec:Preliminaries}
In this section, following Ref.~\cite{JYsingle}, we provide a way to
compute the correlation function of the curvature perturbation from the
dual boundary theory.

\subsection{Wave function}
The cosmological spacetime metric can be given in ADM formalism as
\begin{align}
 & \dd s^2 =  - N^2 \dd t^2 + h_{ij} \left( \dd x^i + N^i \dd t
 \right) \left( \dd x^j + N^j \dd t \right)\,. \label{Exp:metricb}
\end{align}
Here we shall restrict attention to the situation where the metric is
asymptotically de Sitter in the IR and (or) UV.
In the semiclassical picture, this would correspond to the case a period
of slow roll inflation transits to $\Lambda$ domination.

Our starting point is the assumption that the wave function of
the bulk gravitational field is related to the generating functional of
the boundary QFT
\begin{align}
 & \psi_{\rm bulk}[h,\phi] \propto Z_{\rm QFT}[h,\phi]\,, \label{duality}
 \end{align}
where the generating functional $Z_{\rm QFT}$ is given by
\begin{equation}
Z_{\rm QFT} [h,\phi]  =  e^{-W_{\rm QFT}[h,\phi]} =  \int D \chi\,
\exp \left(  - S_{\rm QFT} [\chi,\, h,\, \phi]\right)   \, \label{Exp:Z}.
\end{equation}
(See Ref.~\cite{Maldacena02} and Refs.~\cite{Mata:2012bx, Ghosh:2014kba,
AJ08,AJ09}.) Here $\psi_{\rm
bulk}$ denotes the wave function of the bulk and $\chi$ denotes boundary 
fields, for which the metric $h_{ij}$ and the inflaton $\phi$ act as
sources (the indices in $h_{ij}$ will be omitted when
unnecessary). Since the wave function $\psi_{\rm bulk}$ is complex,
$W_{\rm QFT}$ will have real and imaginary part, and therefore 
the action $S_{\rm QFT}$ cannot be real. It was suggested in
Refs. \cite{AJ08,AJ09} that this local boundary action may actually be
purely imaginary at the fundamental level 
\begin{equation}
S_{QFT}=-iS, 
\end{equation}
with real $S$.  We shall nonetheless stick to the notation in
(\ref{Exp:Z}) because it seems to be widely used, and also to avoid the
proliferation of factors of $i$ in our formal equations, with the
understanding that $S_{QFT}$ is necessarily complex.

\subsection{Correlators in the bulk}
Once we are given the wave function $\psi_{\rm bulk}$, we can compute the
correlators for the bulk. In single field models of inflation, the wave function
of the scalar sector can be expressed by the single gauge invariant variable
$\zeta$ which is the curvature perturbation in the uniform field gauge,
where the inflaton becomes homogeneous. The wave function in the bulk is
then related to the generating functional of the dual quantum field as 
\begin{align}
 & \psi_{\rm bulk}[\zeta] = A\, Z_{\rm QFT}[\zeta] = A\, e^{-  \WQ[\zeta]}\,, \label{Exp:psi}
\end{align}
where we wrote a normalization constant $A$ explicitly.

Using the wave function $\psi_{\rm bulk}[\zeta]$, the probability density function
$P[\zeta]$ is given by 
\begin{align}
 P[\zeta] = \left| \psi_{\rm bulk}[\zeta] \right|^2 
 = |A|^2\, e^{- 2 {\rm Re} \left[\WQ[\zeta]  \right] }\,.
\end{align}
Once we have the partition function $P[\zeta]$, we can calculate the $n$-point
functions for $\zeta$ on the boundary as
\begin{align}
 & \langle \zeta(\bm{x}_1) \zeta(\bm{x}_2)  \cdots \zeta(\bm{x}_n) \rangle = 
  \int D\zeta \, P[\zeta]\,
 \zeta(\bm{x}_1) \zeta(\bm{x}_2)  \cdots \zeta(\bm{x}_n) \,.
\end{align}
Here and hereafter, we abbreviate the argument $t=t(\mu)$ if not necessary. 
The explicit form of the integration measure $D\zeta$ is left unspecified for the time
being. This information is not contained in the boundary QFT, since the
curvature perturbation $\zeta$ is the external field in that context and hence some
additional input may be necessary. Changes in the measure can be usually
represented by local terms in the integrand, which can be incorporated
in a redefinition of $W_{\rm QFT}$. (See also the discussion in
Ref.~\cite{JYsingle}.) We determine the normalization constant $A$, by adopting the normalization condition:
\begin{align}
 & \int D\zeta \, P[\zeta] =1\,. \label{Cond:norm}
\end{align}

Eliminating the background contribution $\WQ[\zeta=0]$ by the
redefinition of $A$, the partition function $P[\zeta]$ is given by 
\begin{align}
 & P[\zeta]=  |A|^2  e^{- \delta W[\zeta] }\,,
\end{align}
where we defined 
\begin{align}
 & \delta W[\zeta] \equiv 2 {\rm Re} \left[ \WQ[\zeta] -
 \WQ[\zeta=0]  \right]\,.  
\end{align}
We expand $\delta W[\zeta]$ as
\begin{align}
 & \delta W[\zeta] = \sum_{n=1}^n \frac{1}{n!} \int \dd^d \bm{x}_1
 \cdots \int \dd^d \bm{x}_n  W^{(n)} (\bm{x}_1,\, \cdots,\,
 \bm{x}_n ) \zeta(\bm{x}_1) \cdots \zeta (\bm{x}_n) \,, \label{Exp:dWn}
\end{align}
where
\begin{align}
 & W^{(n)} (\bm{x}_1,\, \cdots,\,
 \bm{x}_n ) \equiv 2 {\rm Re} \left[ \frac{\delta^n \WQ[\zeta]}{\delta
 \zeta(\bm{x}_1) \cdots \delta \zeta(\bm{x}_n) }
 \Bigg|_{\zeta=0} \right]\,. \label{Def:Wn}
\end{align}
Once we obtain $W^{(n)}(\bm{x}_1,\, \cdots,\, \bm{x}_n)$, we can give
the $n$-point functions, following the Feynman rules~\cite{JYsingle}.
In particular, the two-point function for $\zeta(\bm{x})$ is given by 
\begin{align}
 & \langle \zeta (\bm{x}_1) \zeta (\bm{x}_2) \rangle = W^{(2)\, -1}
 (\bm{x}_1,\, \bm{x}_2) \,, \label{Exp:PP} 
\end{align}
where $W^{(2)\, -1} (\bm{x}_1,\, \bm{x}_2)$ denotes
the inverse matrix of $W^{(2)}(\bm{x}_1,\, \bm{x}_2)$, which satisfies
\begin{align}
 &  \int \dd^d \bm{x}' W^{(2)} (\bm{x}_1,\, \bm{x}') 
 W^{(2)\, -1} (\bm{x}',\, \bm{x}_2) =  \delta(\bm{x}_1 -
 \bm{x}_2 )\,.  \label{Rel:W2W2I}
\end{align}
In this paper, we consider only the tree-level diagrams, neglecting
contributions from loop diagrams, which are suppressed in the large $N$
limit~\cite{JYsingle}. In Ref.~\cite{JYsingle}, it was shown
that the power spectrum and the bi-spectrum computed by using the vertex function 
$W^{(n)}(\bm{x}_1,\, \cdots,\, \bm{x}_n)$ agree with the ones obtained in
Ref.~\cite{BMS} by using the holographic renormalization method.

\section{Deformed conformal field theory}  \label{Sec:dCFT}
In this section, we describe the features of the $d$-dimensional field theory
dual to the $(d+1)$-dimensional inflationary spacetime. For simplicity,
we shall assume that $d$ is odd since in this case a conformal field
theory (CFT) has no conformal anomaly. We consider a local field theory where
the the conformal symmetry is broken by the introduction of a deformation
operator:
\begin{align}
 & S_{\rm QFT} [\chi] = S_{\rm CFT} [\chi] +  \int \dd \Omega_d u  
 O(\bm{x}) \,. \label{Exp:Su}
\end{align}
Here $\dd \Omega_d$ is the $d$-dimensional invariant volume and 
$S_{\rm CFT}$ is the action at the UV or IR fixed point (FP), which preserves the
conformal symmetry, while $u$ is a coupling accompanying the deformation operator $O$. In this section, assuming the flat space, we solve the RG flow. Then, the coupling constant $u$ 
varies depending on the renormalization scale $\mu$. The $\mu$ dependence
of $u$ will be reinterpreted as the time dependence of the background
scalar field in the bulk.

\subsection{Formulas}
Before we solve the RG flow, we summarize the formulas for the CFT in
the flat $\mathbb{R}^d$. The conformal invariance
determines the two-point function and the three point function as
\begin{align}
 & \langle O(\bm{x}) O(\bm{y}) \rangle_{\rm CFT} = \frac{c}{|\bm{x} - \bm{y}|^{2\Delta}} \,,  \label{Eq:2pCFT}
\end{align}
and 
\begin{align}
 & \langle O(\bm{x}) O(\bm{y}) O(\bm{z}) \rangle_{\rm CFT} =
 \frac{C}{|\bm{x} - \bm{y}|^{\Delta}
 |\bm{y} - \bm{z}|^{\Delta} |\bm{z} -
 \bm{x}|^{\Delta}} \,,  \label{Eq:3pCFT}
\end{align}
with the constant parameters $c$ and $C$. Here, $\Delta$ is
the scaling dimension of the operator $O$. The operator product expansion (OPE) is then given by
\begin{align}
 & O(\bm{x}) O(\bm{y}) = \frac{c}{|\bm{x} -
 \bm{y}|^{2\Delta}} +  \frac{C}{c} \frac{O(\bm{x})}{|\bm{x} -
 \bm{y}|^{\Delta}} + \cdots   \label{OPE}
\end{align}
for $|\bm{x} - \bm{y}| \to 0$~\cite{WilsonOPE}. In Eq.~(\ref{OPE}), we abbreviated the
non-singular terms in the limit $|\bm{x} - \bm{y}| \to 0$.

\subsection{RG equation}  \label{Sec:RG}
Following Ref.~\cite{Klebanov11}, we study the RG flow for the local deformed CFT with
the action (\ref{Exp:Su}). The generating functional is given by
\begin{align}
 & Z_{\rm QFT} = \int D\chi
\exp \left( - S_{\rm CFT} - \int \dd^d \bm{x} u O(\bm{x}) \right)\,. \label{Exp:ZdCFT}
\end{align}
First we consider the correlation functions with the UV cutoff scale
$\mu_0$, which are given by
\begin{align}
 & \langle O(\bm{x}_1) \cdots O(\bm{x}_n) \rangle_{\mu_0}
  \cr
&\,= \frac{1}{Z_{\rm QFT}} \int D\chi\,  O(\bm{x}_1) \cdots
 O(\bm{x}_n) \exp \left( - S_{\rm CFT} - \int \dd^d \bm{x} u_0
 O(\bm{x}) \right) \,, \label{Exp:On}
\end{align}
where $u_0$ denotes the bare coupling constants $u$ at $\mu=\mu_0$.
Since we introduced the UV cutoff at $\mu = \mu_0$, all points
$\bm{x}_i$ with $i=1,\, \cdots,\, n$ should be separated with the distance greater than
$1/\mu_0$. The correlation functions for the deformed CFT
can be understood as those for the CFT with the insertion
\begin{align}
 & e^{- \int \dd^d \sbm{x}\, u_0 O(\sbm{x})} \cr
 &\, = 1 - \int \dd^d
 \bm{x}\, u_0 O(\bm{x})  + \frac{1}{2}  \int \dd^d
 \bm{x}\,\int_{|\sbm{x} - \sbm{y}| > 1/\mu_0} \hspace{-2pt} \dd^d \bm{y}\,  u^2_0
 \, O(\bm{x}) O(\bm{y}) + \cdots \,.  \label{expansion}  
\end{align} 
Integrating over the modes between $\mu$ and $\mu_0$ with the aid of the
OPE (\ref{OPE}), we find that the integration of the modes 
$1/\mu_0 \leq |\bm{x} - \bm{y}| \leq 1/\mu$ in the third term of
Eq.~(\ref{expansion}) can be recast into
\begin{align}
 &  \frac{1}{2}  \int \dd^d \bm{x}\,\int_{1/\mu_0 <|\sbm{x} - \sbm{y}| <
 1/\mu} \hspace{-2pt} \dd^d \bm{y}\,  u_0^2\, O(\bm{x})
 O(\bm{y}) \cr
 & = \frac{1}{2} u_0^2\,
 \frac{C}{c}   \int \dd^d \bm{x}\,
 O(\bm{x}) \int_{1/\mu_0 <|\sbm{x} - \sbm{y}| < 1/\mu} \hspace{-2pt}
 \dd^d \bm{y}\, \frac{1}{|\bm{x} - \bm{y}|^{\Delta(\mu_0)}} + \cdots \cr
 & =  -  \frac{1}{2} u^2_0 \tilde{C}
 \frac{\mu^{\lambda} - \mu_0^{\lambda}}{\lambda} \int \dd^d \bm{x}\,
 O(\bm{x}) + \cdots  \,. \label{integration}
\end{align}
Here, we introduced
\begin{align}
 & \lambda \equiv \Delta(\mu_0) - d\,,  \label{Def:lambdaa}
\end{align}
and
\begin{align}
 & \tilde{C} \equiv {\rm Vol}(S^{d-1})\, 
 \frac{C}{c} = \frac{2 \pi^{d/2}}{\Gamma(d/2)}  
 \frac{C}{c} \,,
\end{align}
where $ {\rm Vol}(S^{d-1})$ is the integration of $(d-1)$-dimensional
sphere. Then, the integration of these modes gives rise to the running
of the coupling constant $u$ as 
\begin{align}
 & u(\mu) = u_0 +  \frac{1}{2} u^2_0 \tilde{C}\,
 \frac{\mu^{\lambda} -
 \mu_0^{\lambda}}{\lambda} + {\cal O} (u_0^3)
 \,. \label{Exp:uarun}
\end{align}
In the second and third lines of Eq.~(\ref{integration}), we included
only the terms which contribute to the running of the coupling constants
$u$.

Let us now introduce the dimensionless coupling constants $g$ as
\begin{align}
 & g(\mu) \equiv \mu^{\lambda} u (\mu) \,.
\end{align}
By using Eq.~(\ref{Exp:uarun}), the running of $g(\mu)$ is given by
\begin{align}
 & g(\mu) = g_0 \left( \mu \over \mu_0 \right)^{\lambda}   +
 \frac{1}{2} g^2_0 \tilde{C}
 \frac{  \left( \mu \over \mu_0 \right)^{2 \lambda} -
  \left(\mu \over \mu_0 \right)^{\lambda}}{\lambda} + {\cal
 O} (g_0^3)  \,,  \label{Exp:ga}
\end{align}
where we defined $g_0$ as
\begin{align}
 & g_0 \equiv \mu_0^{\lambda} u_0\,.
\end{align}
Using the dimensionless coupling constant $g(\mu)$, we introduce the
beta function as
\begin{align}
 & \beta(\mu) \equiv \frac{\dd g(\mu)}{\dd \ln \mu} \,. \label{Def:beta}
\end{align}
Inserting Eq.~(\ref{Exp:ga}) into Eq.~(\ref{Def:beta}), we obtain the RG
equation as
\begin{align}
 & \beta(\mu) = \lambda\, g(\mu) + \frac{\tilde{C}}{2} g^2(\mu)  + {\cal O} (g^3)\,.  \label{Eq:betas}
\end{align}
The second term stems from the quantum corrections, which lead to the
deviation from the classical scaling. This analysis is valid for small $g$, in which case the RG flow can be solved
perturbatively. Note that the beta function does not include the UV
cutoff $\mu_0$ explicitly, so we can send it to infinity.

\subsection{Solving RG flow}
In the previous subsection, we obtained the RG equation
(\ref{Eq:betas}). Next, solving the RG equation, we examine the
evolution of $g(\mu)$ more explicitly. In the following, assuming that the dimensionless coupling constant $g(\mu)$
is kept small everywhere along the RG flow, we neglect the terms with
${\cal O}(g^3)$ in Eq.~(\ref{Eq:betas}). Assuming the presence of the IR and UV FPs, we request that in
the vicinity of the UV FP, the operator should be a relevant
one and in the vicinity of the IR FP, should be
an irrelevant one. Without loss of generality, we can
assume $g>0$.

\begin{figure}[t]
\begin{center}
\begin{tabular}{cc}
\includegraphics[width=8cm]{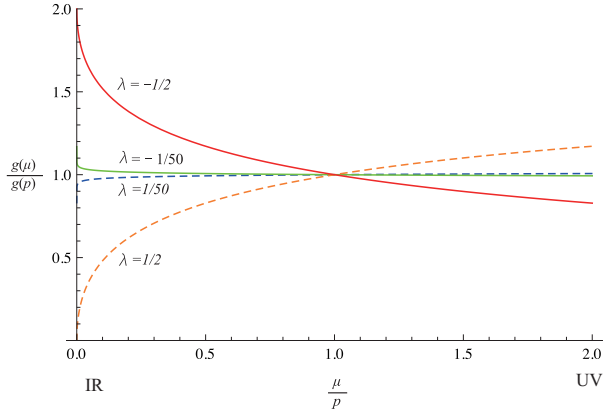}
\end{tabular}
\caption{The evolution of the dimensionless constant $g(\mu)$ for
 $\lambda=1/50, 1/2, -1/2, -1/50$.}  
\label{Fg:gmu}
\end{center}
\end{figure}

Equation (\ref{Eq:betas}) reveals that only when
\begin{align}
 & \frac{\lambda}{\tilde{C}} < 0\,,
\end{align}
the RG flow has two FPs at $g(\mu)=0$ and $g(\mu)= - 2
\lambda/\tilde{C}$. In this case, the RG equation (\ref{Eq:betas}) can be solved as
\begin{align}
 & g(\mu) =  \frac{2}{ 1+ \left( \mu \over p \right)^{\lambda}} 
  \left( \mu \over p \right)^{\lambda} g(p) \,, \label{Sol:gs}
\end{align}
with
\begin{align}
 & g(p) \equiv - \frac{\lambda}{\tilde{C}}\,,  \label{Def:gp}
\end{align}
where a pivot scale $p$ is introduced as an integration constant. 
In the case with $\lambda <0$, the coupling constant $g(\mu)$ flows
from $0$ in the UV to $2 g(p)$ in the IR. On the
other hand, in the case with $\lambda > 0$, the coupling constant
$g(\mu)$ flows from $2 g(p)$ in the UV to $0$ in the
IR. Figure~\ref{Fg:gmu} shows the evolution of $g(\mu)$ for both
positive and negative values of $\lambda$.

\section{Deriving the $\zeta$ correlators}   \label{Sec:vertexfn}
In this section, we consider the boundary QFT in the presence 
of the curvature perturbation $\zeta$, playing the role of an external source. Then, using
the generating functional for the deformed conformal field theory,  
we will derive the relation between the vertex function 
$W^{(n)} (\bm{x}_1,\,\cdots,\, \bm{x}_n )$ and the correlation functions of the boundary
operator $O$ in flat space.

\subsection{Gauge conditions}
In cosmological perturbation theory, the freedom to choose coordinates is usually referred to as gauge
freedom. This corresponds to a choice of the time slicing, and with the choice of spatial coordinates on each slice. 
In the holographic description, we may think of a constant time slice as a holographic plane in which the QFT lives,
while different times correspond to different values of the
renormalization scale. Correlators of the gauge invariant variable
$\zeta$ should be independent of the gauge choice (see also the 
discussion in Ref.~\cite{Maldacena02}).

As the dictionary which relates the bulk and the boundary, in this
paper, we assume that the coupling constant $g$ is related to the
inflaton as 
\begin{align}
 & g(\mu,\, \bm{x}) = \phi(t(\mu),\, \bm{x})\,, \label{Rel:gphi}
\end{align}
where we set $\Mp=1$. Since we assumed that the renormalization scale $\mu$ in the boundary is associated
with the time coordinate $t$ in the bulk, we wrote the time coordinate of
$\phi$ as $t(\mu)$. Equation (\ref{Rel:gphi}) does not provide any
restriction on the bulk dynamics, since we are not specifying the form
of the kinetic term for $\phi$, or its potential. One may be interested
in using more general function $g[\phi]$ instead of the linear one
(\ref{Rel:gphi}), but this can be understood simply as a change of
variable which should not change the physics\footnote{In fact, we can show explicitly that the power spectrum of $\zeta$ is independent of the functional form of the local relation $g(\mu,\, \bm{x}) =
g[\phi(t(\mu),\, \bm{x})]$ (assuming that it is invertible).}.

In the following, we
compute the correlators of $\zeta$, using two gauges. In one gauge,
we choose the holographic plane by requiring
\begin{align}
 & \delta u(t,\, \bm{x})=\delta g(t,\, \bm{x})=\delta \phi(t,\, \bm{x})=0\,,
\end{align}
and the spatial coordinates by requiring that the spatial metric
should be in the form: 
\begin{align}
 & h_{ij} = a^2(t) e^{2\cR(t,\, \sbm{x})} \delta_{ij}\,,  \label{Exp:gauge}
\end{align}
where $\cR$ is the curvature perturbation. In this paper, the tensor
perturbation will be completely neglected. As in the standard CPT, we refer to this gauge
as the uniform field gauge. By definition, the curvature  
perturbation in the uniform field gauge gives the gauge invariant perturbation
$\zeta$, {\it i.e.}, 
\begin{align}
 & \zeta(t,\, \bm{x}) = \cR (t,\, \bm{x}) \big|_{\delta g=0} \,.
\end{align}
In the other gauge, we choose the slicing and the spatial coordinates, requesting
\begin{align}
 & \cR(t,\, \bm{x}) =0\,
\end{align}
and
\begin{align}
 & h_{ij} = a^2(t) \delta_{ij}\,,  \label{Exp:gaugef}
\end{align}
respectively. We refer to this gauge as the flat gauge. In the flat gauge, the scalar perturbation is described solely
by the fluctuation of the coupling constant $\delta g(t,\, \bm{x})$. In
the following, we denote the fluctuation $\delta g(t,\, \bm{x})$ in the
gauge $\cR(t,\, {\bf x})=0$ as
\begin{align}
 & \delta g_f(t,\, {\bm x}) \equiv \delta g (t,\, \bm{x}) \big|_{\cR=0}=
 \delta \phi(t,\, \bm{x}) \big|_{\cR=0}\,,
\end{align}
which is also gauge invariant.

In the standard cosmological perturbation theory, performing the gauge
transformation, we find that the curvature perturbation in the uniform
field gauge, $\zeta$, is related to the fluctuation of the inflaton in the flat gauge
$\delta g_f$ as (see e.g. \cite{Maldacena02, Urakawa:2010kr})
\begin{align}
  \zeta = -\frac{H}{\dot{\phi}} \delta g_f + \frac{\varepsilon_2}{4}
   \left( \frac{H}{\dot{\phi}} \right)^2 \delta g_f^2  
   + \cdots \,. \label{Rel:zetadphi}
\end{align}
Here we abbreviated the sub-leading terms at large scales, as well as
higher orders in $\delta g_f$, and we used
the horizon flow functions, defined as 
\begin{eqnarray}
  \varepsilon_n \equiv \frac{1}{\varepsilon_{n-1}} \frac{\dd}{\dd \ln a} \varepsilon_{n-1}
\end{eqnarray}
for $n \geq 1$, with
\begin{eqnarray}
  \varepsilon_1 \equiv  \frac{1}{2} \frac{\dot{\phi}^2}{H^2}\,. \label{Def:SR1}
\end{eqnarray}
Notice that for the scalar field with the non-canonical kinetic term,
our $\varepsilon_1$ does not coincide with the standard definition of
the horizon flow functions, given by $\varepsilon_1= - \dot{H}/H^2$.

\subsection{Renormalization and counterterms}
To derive finite correlation functions, we need to perform
renormalization. The studies based on the holographic renormalization provide
the necessary counterterms and renormalized action in the bulk (see,
for instance, Refs.~\cite{Verlinde, de Haro:2000xn, Kostas_RG, Kostas_RG02, Kiritsis:2014kua}). Meanwhile, to derive the renormalized
correlation functions based on the boundary computation, we need to introduce the
counterterms and determine the renormalized action in the boundary
theory. One may expect that the introduction of the counterterms will
alter the boundary theory through the contributions of the following three different types:
\begin{enumerate}
 \item Additive local contributions which are independent of the boundary
       operators. These are made out of invariants constructed from the sources.
 \item Contributions proportional to the boundary operator $O$.
 \item Contributions proportional to new operators $O_a$ with
       $a=1,\,2,\, \cdots$.
\end{enumerate}
Here, we consider the counter terms which keep the boundary theory
local. Contributions of the second type will correspond to the renormalization
of the coupling constant $u$ or $g$. Contributions of the third type
will take us to the multi-field case. In this paper, we disregard
this possibility, concentrating in the one field case.

Choosing the uniform field gauge with $\delta g=0$, the
generating functional for the boundary theory is expressed as  
\begin{align}
 & Z_{\rm QFT}[\zeta]  = \int D\chi\,
 e^{ - S_{\rm QFT}[\zeta,\,\chi]} \label{Exp:ZdCFTfull}
\end{align}
with the action for the boundary QFT given by
\begin{align}
 & S_{\rm QFT}[\zeta,\, \chi] \equiv S_{\rm CFT}  
 -  \int \dd^d \bm{x} e^{d \zeta(\sbm{x})} g(\mu) {\cal O}(\mu,\,
 \bm{x}) + S_{\rm source}[\zeta]. \label{Exp:Sren}
\end{align}
For notational convenience, we introduced
\begin{align}
 & {\cal O}(\mu,\,\bm{x}) \equiv \mu^{-\lambda} O(\mu,\,\bm{x})\,.
\end{align}
In the right-hand side of the action, we absorbed the scale factor which appears from the
invariant volume into $g(\mu)$. The third term in Eq.~(\ref{Exp:Sren})
denotes the additive contributions of the first type. Note that this
term can be factorized in the generating functional $W_{\rm QFT}=-\ln
Z_{\rm QFT}[\zeta]$ as 
\begin{align}
 & W_{\rm QFT}[\zeta] = - \ln \left[ \int D\chi e^{- S_{\rm CFT} - \int \dd^d \sbm{x} e^{d
 \zeta} g {\cal O}}  \right] + S_{\rm source} [\zeta] \,. \label{Exp:WQFTR}
\end{align} 
Therefore, when we derive the vertex function $W^{(n)}$ by taking the
derivative with respect to $\zeta$ as in Eq.~(\ref{Def:Wn}), the derivative
which operates on $S_{\rm source}[\zeta]$ gives only the 
disconnected product of the ultralocal term where all arguments $\bm{x}_i$s
coincide and the correlators of ${\cal O}$ derived from the first term
in the right hand side of Eq.~(\ref{Exp:WQFTR}). Since we are interested in 
connected diagrams, we focus on the contribution from the first term.

\subsection{$\zeta$ correlators from gauge invariance}
\label{SSec:WnCP}
In this subsection, we will study the relation between the vertex
function $W^{(n)}$ and the correlators of ${\cal O}$, focusing on the fact that in the single field model
where the gauge invariance is preserved, the gauge-invariant variable
$\zeta$ can be expressed as a functional of the gauge invariant variable
$\delta g_f$, {\it i.e.}, $\zeta=\zeta[\delta g_f]$ or inversely
$\delta g_f= \delta g_f[\zeta]$. For the time being, we proceed our
discussion without invoking the explicit form of $\zeta=\zeta[\delta
g_f]$. Equation (\ref{Def:Wn}) states that the vertex function
$W^{(n)}(\bm{x}_1,\, \cdots,\, \bm{x}_n)$ is given by the $n$-th
derivative of the generating functional $W_{\rm QFT}= - \ln Z_{\rm
QFT}$. Recasting the derivative with respect to $\zeta$ into the
derivative with respect to $\delta g_f$ by using the schematic relation $\delta g_f= \delta
g_f[\zeta]$, we obtain 
\begin{align}
 &  \frac{\delta W_{\rm
 QFT}}{ \delta \zeta(\bm{x})} = \int \dd^d \bm{y} \frac{\delta g_f(\bm{y})}{\delta \zeta (\bm{x})} \frac{\delta W_{\rm
 QFT}}{ \delta g_f(\bm{y})} \,.  \label{WTcp}
\end{align}
When $\delta g_f(\bm{x})$ is locally related to $\zeta(\bm{x})$ as in
the large scale limit as given in Eq.~(\ref{Rel:zetadphi}), using
Eq.~(\ref{Rel:gphi}), we find that
$\delta g_f(\bm{x})$ is also related to $\zeta(\bm{x})$ locally. Then, we can rewrite Eq.~(\ref{WTcp}) as
\begin{align}
 &  \frac{\delta W_{\rm
 QFT}}{ \delta \zeta(\bm{x})} =  -B(\bm{x}) \frac{\delta W_{\rm
 QFT}}{ \delta g_f(\bm{x})} \,,  \label{WTcp2}
\end{align}
where we introduced $B(\bm{x})$ as
\begin{align}
 & B(\bm{x}) \equiv - \frac{\partial \delta g_f(\bm{x})}{\partial \zeta
 (\bm{x})} \,.
\end{align}
Note that Eq.~(\ref{WTcp2}) states that taking the derivative
with respect to the gauge invariant variable $\zeta$ is equivalent to taking the
derivative with respect to $\delta g_f$ up to the factor $-B(\bm{x})$.

Next, using Eq.~(\ref{WTcp2}), we derive the relation between the
vertex function $W^{(n)}$ and the correlators of ${\cal O}$ in the flat
space. Taking the derivative of Eq.~(\ref{WTcp2}) with respect to $\zeta(\bm{x})$, we
obtain 
\begin{align}
 & \frac{\delta^2 W_{\rm
 QFT}}{ \delta \zeta(\bm{x}_1) \delta \zeta(\bm{x}_2)} =
 B(\bm{x}_1) B(\bm{x}_2)  \frac{\delta^2 W_{\rm
 QFT}}{ \delta g_f(\bm{x}_1)  \delta g_f(\bm{x}_2)} - \frac{\delta
 B(\bm{x}_1)}{\delta \zeta(\bm{x}_2)} \frac{\delta W_{\rm
 QFT}}{\delta g_f(\bm{x}_1)}\,. \label{W2cp}
\end{align}
Using Eqs.~(\ref{Def:Wn}), (\ref{W2cp}), and
\begin{align}
  \frac{\delta^n W_{\rm
 QFT}}{ \delta g_f(\bm{x}_1) \cdots  \delta g_f(\bm{x}_n)}
 \bigg|_{\delta g_f=0} = (-1)^{n+1} \langle {\cal O}(\bm{x}_1) \cdots
 {\cal O} (\bm{x}_n) \rangle_\mu \,,
\end{align}
we obtain  
\begin{align}
 W^{(2)}(\bm{x}_1,\, \bm{x}_2) & = - 2 {\rm Re} \left[ B_1^2  \langle {\cal
 O}(\bm{x}_1) {\cal O}(\bm{x}_2)  \rangle_\mu \right]  \,,
\end{align}
where we set the renormalization condition such that the ultralocal term
with $\delta(\bm{x}_1 - \bm{x}_2)$ is canceled by the contribution from
the quadratic term in $S_{\rm source}[\zeta]$. Here, introducing the
$\mu$ dependent function $B_n(\mu)$ as 
\begin{align}
 B_n(\mu) \equiv - \frac{\partial^n \delta g_f(\mu,\, \bm{x})}{\partial
 \zeta^n (\mu,\, \bm{x})}\bigg|_{\zeta=0} \,,
\end{align}
we expressed $B(\bm{x})$ as
\begin{align}
 & B(\bm{x}) \big|_{\zeta=0} =B_1(\mu)\,.
\end{align}
The vertex function $W^{(n)}$ with $n\geq 3$ can be obtained similarly
and we find that $W^{(n)}$ is given in the form 
\begin{align}
 &W^{(n)}(\bm{x}_1,\, \cdots,\, \bm{x}_n) \cr
&= -2 {\rm Re}
 \Bigl[ B_1^n  \langle {\cal O}(\bm{x}_1) \cdots   
 {\cal O}(\bm{x}_n) \rangle_\mu \cr & \qquad \qquad \,\,
  +  B_2  B_1^{n-2} \left\{ \delta (\bm{x}_1 - \bm{x}_2)  \langle {\cal O}(\bm{x}_2) \cdots  
 {\cal O}(\bm{x}_n) \rangle_\mu  + \left( \cp \right)  \right\} + \cdots \cr
 & \qquad \qquad\,\,    + \! \sum_{m=1}^{[n/2]} \! B_m B_{n-m}
 \bigl\{ \delta(\bm{x}_1 - \bm{x}_2) \cdots
 \delta(\bm{x}_{m-1}-\bm{x}_m) \delta(\bm{x}_{m+1} - \bm{x}_{m+2}) \cdots
 \delta(\bm{x}_{n-1} - \bm{x}_n) \cr
 & \qquad \qquad  \qquad \qquad \qquad \qquad \qquad  \times \langle {\cal O}(\bm{x}_m) {\cal O}(\bm{x}_{m+1}) \rangle_\mu
 + \left( \cp \right) \bigr\}  \Bigr] \,, \label{Exp:Z0gn}
\end{align}
where $[x]$ denotes the Gauss's floor function. Here, the delta functions
appeared by taking the derivative of $B(\bm{x})$ with respect to
$\zeta(\bm{x})$, for instance, as
$$
 \frac{\delta B(\bm{x}_1)}{\delta \zeta (\bm{x}_2)} \bigg|_{\zeta=0} =
\delta(\bm{x}_1 - \bm{x}_2) B_2(\mu)\,.
$$
In Eq.~(\ref{Exp:Z0gn}),
we again eliminated the ultralocal term, using the contribution from the
$n$-th term in $S_{\rm source}[\zeta]$. Once the relation between $\delta \phi$ and
$\zeta$ is given, using Eq.~(\ref{Exp:Z0gn}), we can express the vertex function $W^{(n)}$ by
$B_n$ and the correlators of ${\cal O}$ in the flat space. Namely, when
we use the relation (\ref{Rel:zetadphi}), we can express $B_n$ as
\begin{align}
 & B_1 = \frac{\dot{\phi}}{H} = \frac{\dd \phi}{\dd \ln a} \,, \label{Exp:B1cp} \\ 
 & B_2 = -  \frac{\dot{\phi}}{H}  \frac{\varepsilon_2}{2} = - 
  \frac{\dd B_1}{\dd \ln a}  = -  \frac{\dd^2 \phi}{\dd \ln a^2}  \,.  \label{Exp:B2cp}
\end{align} 
Thus, the relation between the vertex function $W^{(n)}$ and the correlators of
${\cal O}$ is specified by invoking the relation between
$\zeta$ and $\delta g_f$, derived by performing the gauge
transformation in the cosmological perturbation theory. In Appendix
\ref{Sec:WTidentity}, we seek for an alternative way to relate
$W^{(n)}$ to the correlators of ${\cal O}$. The ambiguity discussed in
Appendix \ref{SSec:ambiguity} can be eliminated by using the relation
(\ref{Rel:zetadphi}).

\section{Conservation of the curvature perturbation $\zeta$}  \label{Sec:conservation}
In this section, after we overview the discussion about the conservation
of the curvature perturbation $\zeta$ based on the standard CPT, we
address the conservation of the curvature perturbation from holography.

\subsection{Conservation in the standard cosmological perturbation theory}
In cosmological perturbation theory (CPT) the conservation of the curvature perturbation $\zeta$ holds in the
large scale limit for the adiabatic time evolution, when the matter content is dominated by
a single species~ \cite{Weinberg, WMLL, MW03, LMS, LV05}. This is useful, for instance, in order to evolve the predicted
distribution function for $\zeta$ through the process of reheating, the details of which are largely unknown. 
For a barotropic perfect fluid, the conservation of $\zeta$ at large
scales can be derived directly from the conservation of the energy
momentum tensor, without invoking the theory of gravity. In turn, the conservation of the energy
momentum tensor in a local theory just relies on the equivalence
principle, usually implemented through general covariance.  

Since the validity of the conservation of $\zeta$ is so generic, any
evidence that conservation is violated can provide some useful insight
on the nature of the underlying theory of inflation. In the following
subsection, we will study the validity of the conservation of $\zeta$
based on the boundary computation. Before that, let us briefly summarize
the discussion based on the standard CPT. (The conservation of $\zeta$
has been mostly discussed for $d=3$, so in this section, we consider
$d=3$. However, it will hold also in other dimensions.)

In Ref.~\cite{WMLL}, Wands {\it et al.} showed the conservation of
$\zeta$ for the barotropic fluid, whose  energy density perturbation
$\delta \rho$ is proportional to the pressure perturbation $\delta p$,
at linear order in perturbation. Using the energy conservation equation $n^\nu
T^\mu\!_{\nu;\mu}=0$ with the unit timelike vector $n^\nu$: 
\begin{align}
 & \delta \dot{\rho} = - 3 H (\delta \rho + \delta p) - 3 \dot{{\cal R}}
 (\rho + p) + O \left ( (k/aH)^2 \right)\,, 
 \end{align}
where $\rho$ and $p$ are the background values of the energy density and
the pressure, they found that in the uniform density gauge $\delta
\rho=0$, the curvature perturbation is conserved in time at large scales.  This argument was extended to the
non-linear order by Lyth {\it et al.} based on the gradient expansion in
Ref.~\cite{LMS} and also by Langlois and Vernizzi based on the covariant
approach in Ref.~\cite{LV05}.  Their arguments proceed independent of
the theory of gravitation in the bulk.

Note that when we consider the scalar field, the adiabatic condition,
that states the pressure is expressed only by the energy density,
becomes less clear. In fact, even in the single field case, the adiabatic condition is not
necessarily satisfied. In Ref.~\cite{NS}, Naruko and Sasaki considered the Galileon-type scalar
field whose equation of motion involves up to the first derivative of the
metric. Then, using the equation of motion with the aid of the gradient
expansion, they showed that in the attractor phase, during which the
equation of motion for $\phi$ becomes the first order as $\dot{\phi} =
f(\phi)$, the uniform field gauge can be chosen and the curvature
perturbation in this gauge, $\zeta$, is conserved at large scales.  (See
also Ref.~\cite{Gao:2011mz}.) So far, we have summarized the
discussions about the classical evolution which does not include the
loop corrections. The conservation of $\zeta$ is discussed also in the presence of loop
corrections~\cite{ABG, SZ}, but the validity of the conservation for the loop
corrections is still unclear~\cite{SRV2, IRreview}.  

\subsection{Conservation from holography?}
In this subsection, we investigate the conservation of the
gauge-invariant curvature perturbation $\zeta$ based on holography. 
In holography, the conservation of $\zeta$ can be addressed by studying
the $\mu$ dependence, which is interpreted as the time dependence in
the cosmological evolution. Notice that since the renormalized
theory, obtained after integrating out the UV modes, consists only the
wavenumber $k$ with $k < \mu$, the correlators of $\zeta$ given by the
renormalized boundary theory will describe the evolution of
$\zeta$ at large scales.

Here, we note that the $n$-point function of $\zeta$ is described solely in terms of the vertex functions
$W^{(m)}$ with $m \leq n$. For instance, as is shown in
Ref.~\cite{JYsingle}, the power spectrum of $\zeta$ is given by
\begin{align}
 & \langle \zeta(\bm{x}_1) \zeta(\bm{x}_2) \rangle_{\rm conn}  =
 \WI (\bm{x}_1,\, \bm{x}_2)\,, \label{Exp:2pzeta}
\end{align}
with the inverse matrix of $W^{(2)}(\bm{x}_1,\, \bm{x}_2)$, and the bi-spectrum is given by 
\begin{align}
 & \langle \zeta(\bm{x}_1) \zeta(\bm{x}_2) \zeta(\bm{x}_3) \rangle_{\rm conn} = - \int
 \prod_{i=1}^3  \dd^d \bm{y}_i\, \WI(\bm{x}_i\,,\bm{y}_i)\,
 W^{(3)} (\bm{y}_1,\,\bm{y}_2,\,\bm{y}_3)\,.   \label{Exp:BSzeta}
\end{align}  
To make the power spectrum of $\zeta$ conserved, the vertex
function $W^{(2)}$ should be independent of $\mu$. Given that the power
spectrum is conserved, to further make the bi-spectrum of $\zeta$ conserved, the vertex
function $W^{(3)}$ should be also independent of $\mu$. Thus, to make
all the $m$-point functions of $\zeta$ with $m\leq n$ conserved, the 
vertex functions $W^{(m)}$ with $m\leq n$ should be totally independent of $\mu$.
Therefore, in the following, we study the $\mu$ dependence of the vertex
function $W^{(n)}$.

In this subsection, we study whether the correlators of $\zeta$ become $\mu$ independent or
not under the following two assumptions: 
\begin{itemize}
 \item The gauge invariant variables $\zeta$ and $\delta g_f$ are
       locally related schematically as
\begin{align}
 & \zeta(\bm{x}) = \zeta[\delta g_f (\bm{x})]\,. \label{ASM:local}
\end{align} 
 \item The dual boundary theory can be renormalized by using the wave
       function renormalization $Z(\mu)$ as
\begin{align}
 & Z^{-n/2}(\mu) \langle {\cal O}(\bm{x}_1) \cdots {\cal O}(\bm{x}_n) \rangle_\mu =
 Z^{-n/2}(\mu_0) \langle {\cal O}(\bm{x}_1)
 \cdots {\cal O}(\bm{x}_n) \rangle_{\mu_0}\,. \label{ASM:ren}
\end{align}
\end{itemize}
The first assumption will hold generally at large scales (see for
instance Eq.~(\ref{Rel:zetadphi})), unless a 
non-local operator, which typically gives rise to the singular pole
in the limit $k \to 0$, shows up in the relation between $\zeta$ and
$\delta g_f$.

In Sec.~\ref{SSec:WnCP}, using Eq.~(\ref{ASM:local}), we derived the
vertex function $W^{(n)}$ as in Eq.~(\ref{Exp:Z0gn}). First, we consider the power spectrum, given by
the inverse matrix of 
\begin{align}
 W^{(2)}(\bm{x}_1,\, \bm{x}_2) & = - 2 {\rm Re} \left[ B_1^2(\mu)  \langle {\cal
 O}(\bm{x}_1) {\cal O}(\bm{x}_2)  \rangle_\mu \right]  \,. \label{Exp:W2CP}
\end{align}
Inserting Eq.~(\ref{ASM:ren}) into Eq.~(\ref{Exp:W2CP}), we find that
to make $W^{(2)}$ independent of $\mu$, the wave function renormalization $Z(\mu)$ should satisfy
\begin{align}
 & \frac{\dd}{\dd \mu} \left[  B_1(\mu) \sqrt{Z(\mu)} \right] = 0\,.
 \label{Cond:C2p} 
\end{align}
Next, we consider the bi-spectrum of $\zeta$, expressed by $\WI$ and
\begin{align}
 &W^{(3)}(\bm{x}_1,\, \bm{x}_2,\, \bm{x}_3) \cr
&= -2 {\rm Re}
 \Bigl[ B_1^3(\mu)  \langle {\cal O}(\bm{x}_1) {\cal O}(\bm{x}_2) 
 {\cal O}(\bm{x}_3) \rangle_\mu \cr & \qquad \quad \qquad
  +  B_2(\mu)  B_1(\mu) \left\{ \delta (\bm{x}_1 - \bm{x}_2)  \langle {\cal O}(\bm{x}_2)  
 {\cal O}(\bm{x}_3) \rangle_\mu  + \left( \cp \right)  \right\}  \Bigr] \,. \label{Exp:W3cp}
\end{align}
When the condition (\ref{Cond:C2p}) is fulfilled, the first term in the
right-hand side of Eq.~(\ref{Exp:W3cp}) becomes $\mu$ independent. In
addition, to make the terms in the third line of Eq.~(\ref{Exp:W3cp})
independent of $\mu$, $B_2(\mu)$ should be given as
\begin{align}
 & B_2(\mu) = s_2 B_1(\mu)  \label{Cond:B2}
\end{align}
with a constant parameter $s_2$.  Note that using Eq.~(\ref{Exp:B2cp}), we can express the parameter $s_2$ as
\begin{align}
 & s_2 = - \frac{\dd}{\dd \ln a}  \ln B_1  \,.
 \end{align}
Repeating a similar argument, we find that
only if the condition (\ref{Cond:C2p}) is satisfied and $B_m(\mu)$ with
$m\leq n$ is given as 
\begin{align}
 & B_m(\mu) = s_m B_1(\mu)    \label{Cond:B}
\end{align}
with a constant parameter $s_m$, the vertex function $W^{(n)}$ becomes
independent of $\mu$, implying the conservation of $\zeta$.

Next, we examine whether the conditions (\ref{Cond:C2p}) and
(\ref{Cond:B}) can be fulfilled, solving the RG flow explicitly. In
Appendix \ref{Sec:On}, following Ref.~\cite{BMS}, we computed the
renormalized correlators of ${\cal O}$ and then
the wave function renormalization is given as
\begin{align}
 & \sqrt{Z(\mu)} = \mu^{- \lambda} \left[ 1 +  \Bigl(
 \frac{\mu}{p}\Bigr)^{\lambda}  \right]^2 = 4 p^{-\lambda}
 \frac{\beta(p)}{\beta(\mu)} \,. \label{Exp:Zmu}
\end{align}
(The wave function renormalization is discussed from a different
perspective in Ref.~\cite{McFadden:2013ria}.) 
On the second equality, we noted that using Eqs.~(\ref{Def:beta}) and
(\ref{Sol:gs}), the beta function is given as 
\begin{align}
 & \beta(\mu) = 
\frac{4}{\Bigl[1+ \left(\mu \over p
 \right)^{\lambda} \Bigr]^2} \left(\mu \over p \right)^{\lambda}
  \beta(p) = \frac{ \lambda}{1+ \left(\mu \over p
 \right)^{\lambda} } g(\mu) 
\,, \label{Sol:beta1}
\end{align}
with
\begin{align}
 & \beta(p) \equiv \frac{\lambda}{2} g(p)\,.
\end{align}
Using Eq.~(\ref{Exp:Zmu}), we find that the condition (\ref{Cond:C2p}) implies 
\begin{align}
 & B_1(\mu) = {\cal C}  \beta(\mu)\,,   \label{Cond:C2pbeta}
\end{align}
where ${\cal C}$ is a constant parameter. When we use the relation (\ref{Rel:zetadphi}) derived by performing the gauge transformation, the $\mu$ dependent functions $B_1$ and $B_2$ are
given as in Eqs.~(\ref{Exp:B1cp}) and (\ref{Exp:B2cp}). Using
Eqs.~(\ref{Def:beta}), (\ref{Exp:B1cp}), and (\ref{Cond:C2pbeta}), we
find that the renormalization scale $\mu$ should be related to the time
coordinate in cosmology as 
\begin{align}
 &  \ln (\mu/\mu_0) =   {\cal C}  \ln (a/a_0)    \,,  \label{Rel:mua}  
\end{align}  
where $a_0$ denotes the scale factor at the time associated with $\mu_0$. When the RG flow has the conformal FP, since $\mu$ is proportional to $a$ near the FP~\cite{BMS, AJ08, AJ09, Alex11}, ${\cal C}$ should be set to 1.

On the other hand, we find that in general the second condition
(\ref{Cond:B}) cannot be fulfilled. In fact, using
Eqs.~(\ref{Cond:C2pbeta}) and (\ref{Rel:mua}), we obtain 
\begin{align}
 & s_2 = - {\cal C} \frac{\dd}{\dd \ln \mu}  \ln \beta \,,  \label{Exp:s22}
\end{align} 
which can be solved as
\begin{align}
 & \beta(\mu) = \beta_0  \left(\mu \over \mu_0
 \right)^{\lambda} 
\,,   \label{tempbeta}
\end{align} 
with the constant $\lambda=\Delta -d$ given by
\begin{align}
 & \lambda = - \frac{s_2}{{\cal C}} \,.  \label{Exp:lambda}
\end{align}
Therefore, unless we consider the RG flow which has the constant scaling
dimension as in Eq.~(\ref{tempbeta}), the condition (\ref{Cond:C2pbeta}) cannot be
fulfilled along the entire RG flow. It is clear that the beta
function for the RG flow with the two FPs (\ref{Sol:beta1}) deviates from
Eq.~(\ref{tempbeta}) once away from the FPs and then we find
Eq.~(\ref{Cond:C2pbeta}) cannot be satisfied except for the vicinities
of the IR and UV FPs. When the right hand side of Eq.~(\ref{Exp:s22})
ceases to be constant, the bi-spectrum of $\zeta$, namely the terms in
the last line of Eq.~(\ref{Exp:W3cp}), can evolve even at large scales, contradicting the prediction of
the CPT for the single field models of inflation.

Several comments are in order regarding the RG flow whose beta function
is given by Eq.~(\ref{tempbeta}). This RG flow has at most one FP: not
having any FP or having only one FP either in the IR or UV. Using
Eq.~(\ref{Rel:mua}), we can rewrite Eq.~(\ref{tempbeta}) in terms of the
cosmological quantities as 
\begin{align}
 & \beta(a) = \frac{1}{{\cal C}} \frac{\dd \phi}{\dd \ln a}=
 \beta_0 \left(a \over a_0 \right)^{- s_2}\,.  \label{tempbetaC}
\end{align} 
Notice that Eq.~(\ref{tempbetaC}) restricts the evolution of $\phi(a)$
and also the potential $V(\phi)$, indicating that the conservation of
$\zeta$ can be verified only for the specific single field model of
inflation.

When $\beta(\mu)$ is given by Eq.~(\ref{tempbeta}), $\beta(\mu)$ yields
the RG equation (\ref{Eq:betas}) with $\lambda=- s_2/{\cal C}$ and $C=0$. In this case,
the two-point function of ${\cal O}$ is simply given by
\begin{align}
 & \langle {\cal O}(\bm{x}_1) {\cal O}(\bm{x}_2) \rangle_\mu 
 = \left( \mu \over \mu_0 \right)^{-2 \lambda} \frac{c_0}{|\bm{x}_1 -
 \bm{x}_2|^{2(\lambda+ 3)}} \,,  \label{tempO}
\end{align}
where $c_0$ is a constant parameter. Inserting Eqs.~(\ref{tempbeta}) and (\ref{tempO}) into
Eq.~(\ref{Exp:W2CP}), we obtain the vertex function $W^{(2)}$ as
\begin{align}
 & W^{(2)}(\bm{x}_1,\, \bm{x}_2) = - \frac{2 {\cal C}^2 \beta^2_0 c_0 }{|\bm{x}_1 -
 \bm{x}_2|^{2(\lambda+ 3)}}\,.  \label{tempW2}
\end{align}
Using Eqs.~(\ref{Exp:2pzeta}) and (\ref{tempW2}), we obtain the power
spectrum of $\zeta$ as
\begin{align}
 & P_\zeta(k) = - \frac{6}{\pi^2} \frac{1}{{\cal C}^2 \beta^2_0 c_0}
 \frac{1}{k^{3 + 2 \lambda}}\,. \label{Eq:Pkconserved}
\end{align}
Thus, the amplitude and the spectral tilt of $\zeta$ are expressed by
$\beta_0$, ${\cal C}$, and $\lambda=- s_2/{\cal C}$. Notice that
Eq. (\ref{Eq:Pkconserved}) states that to provide a red-tilted spectrum
which is consistent with the observation of the cosmic microwave
background, $\lambda=- s_2/{\cal C}$ should be positive. In that case, the beta
function blows up in the UV, causing the breakdown of conformal
perturbation theory at sufficiently short distances.

\section{Discussion}
In this paper, we examined whether the curvature perturbation 
$\zeta$ is conserved under the change of $\mu$ in a local boundary theory with
the action (\ref{Exp:Sren}). This corresponds to a generic CFT with a
single deformation operator. 

In order to relate the boundary correlators to the correlators of the
gauge invariant curvature perturbation $\zeta$, we have assumed that
$\zeta$ is locally related to the scalar field perturbation in the flat
gauge, $\delta\phi_f$ (or equivalently the perturbation of the coupling
$\delta g_f$) as in Eq.~(\ref{ASM:local}). This assumption certainly
holds in standard cosmological perturbation theory at large scales.  
Also, we have assumed that the boundary operator ${\cal O}$ is
renormalized multiplicatively as in Eq.~(\ref{ASM:ren}). Solving the RG
flow, we found that the power spectrum of $\zeta$ is conserved if we
identify the renormalization scale $\mu$ with the scale factor $a$ in
the bulk, as in Eq.~(\ref{Rel:mua}). But then, it follows that the
bi-spectrum of $\zeta$ cannot be conserved along the entire RG flow. The
only exception is the particular case where the scaling dimension is
constant, as given in Eq.~(\ref{tempbeta}). This special case leads to
an exact power law spectrum of the form (\ref{Eq:Pkconserved}).

In order to have conservation of $\zeta$ along  a generic RG flow, we
need to abandon at least one of the following three  assumptions: the
local relation Eq.~(\ref{ASM:local}), the multiplicative renormalization
Eq.~(\ref{ASM:ren}) or the assumption of a local boundary theory with
the single deformation operator.

The local relation Eq.~(\ref{ASM:local}) follows from
Eqs.~(\ref{Rel:gphi}) and (\ref{Rel:zetadphi}). Instead of imposing
Eq.~(\ref{Rel:gphi}), one may need to seek for a more non-trivial
relation between $g$ and $\phi$ (this issue has been discussed in
AdS/CFT. See, for instance, Ref.~\cite{Megias:2014iwa}). A simple
generalization to the local function $g(\mu,\, \bm{x})=g[\phi(t(\mu),\,
\bm{x})]$ is just a change of variable describing the field, and will
not help to preserve the conservation. The second and third assumptions
are concerned with renormalization. In this paper, we assumed that the
boundary QFT can be renormalized by introducing the counterterm and the
wave function renormalization as in Eqs.~(\ref{Exp:Sren}) and
(\ref{ASM:ren}), respectively. Nevertheless, in general, one may need to
introduce more than one deformation operator to perform the
renormalization. In addition, the QFT may become non-local after the
renormalization. These cases were not addressed in the present
paper. When the boundary theory contains more than one deformation
operator, the corresponding cosmological evolution will be governed by
several scalar fields. Notice, however, that in this case the standard
CPT does not predict the conservation of $\zeta$ any longer. The
possibility of generalizing the duality to the case of a non-local
boundary theory is at present not very well understood. In particular,
it is not clear whether the locality (non-locality) in one side of the
duality implies the locality (non-locality) on the other side. A
relevant discussion can be found in Ref.~\cite{Sundrum:2011ic}, but to
our knowledge this issue has not been fully resolved, at least in the
case when the deviation from dS spacetime becomes important (even in the large $N$ limit). 
If a non-local boundary theory can be dual to a local bulk theory with a single field,
the conservation of $\zeta$ should be predicted also from the boundary
computation. We leave this issue for future research.

In this paper, we discussed the asymptotically dS spacetimes  and the dual
boundary QFT. Since the asymptotically dS spacetimes which we have considered can be transformed
into the asymptotically AdS spacetime by  analytic continuation, the analog of 
$\zeta$ will be conserved along the holographic direction also in the case of asymptotically
AdS spacetimes. Possible implications of our results in this broader context are 
currently under investigation.

\acknowledgments
We would like to thank K.~Skenderis for his valuable comments and early
collaboration. Y.~U. would like to thank S.~Sibiryakov for his
suggestion about the possible origin of the non-conservation. J.~G. and
Y.~U. are partially supported by MEC FPA2010-20807-C02-02, AGAUR
2009-SGR-168, and CPAN CSD2007-00042 Consolider-Ingenio 2010. Y.~U. is
supported by the JSPS under Contact No.~21244033.

\appendix

\section{Correlators of the boundary operator $O$}  \label{Sec:On}
In this section, we compute the renormalized $n$-point functions of $O$. Following
Ref.~\cite{BMS}, we perform the renormalization in the conventional
way, introducing the UV cutoff. Expanding the $n$-point function regarding the deformation term
$\int \dd^d \bm{x} u O$ as
\begin{align}
 & \langle O(\bm{x}_1) \cdots O(\bm{x}_n) \rangle_\mu \cr
 &= \sum_{m=0}^\infty \frac{1}{m!} (- u)^m \int \dd^d \bm{y}_1 \cdots
 \int \dd^d \bm{y}_m \langle O(\bm{x}_1) \cdots
 O(\bm{x}_n) O(\bm{y}_1) \cdots O(\bm{y}_m) \rangle_{\mu,\, 0} \,,
\end{align}
we express the $n$-point function as the summation of the correlators
for the ``CFT'' with the cutoff at $\mu$ (here we put the quotes
since the theory cannot be the exact CFT due to the presence of the
cutoff), which we denote as 
$\langle \cdots \rangle_{\mu,\,0}$. We first compute 
\begin{align}
 & I_m^{(n)} (\bm{x}_1,\, \cdots,\, \bm{x}_n,\, \mu) \cr
 & \equiv \int \dd^d \bm{y}_1 \cdots
 \int \dd^d \bm{y}_m \langle O(\bm{x}_1) \cdots
 O(\bm{x}_n) O(\bm{y}_1) \cdots O(\bm{y}_m) \rangle_{\mu,\, 0} \,.  \label{Def:Imn}
\end{align}
Note that since this is the correlators with the cutoff $\mu$, all the
points $\bm{z}$ and $\bm{z}'$ which are either $\bm{x}_i$ with $i=1, \cdots,\, n$ or
$\bm{y}_i$ with $i=1, \cdots,\, m$ should satisfy
\begin{align}
 & |\bm{z} - \bm{z}'| \geq 1/\mu\,.  \label{Cond:cutoff}
\end{align}
First, we compute $I^{(n)}_1$, which is given as
\begin{align}
 & I_1^{(n)} (\bm{x}_1,\, \cdots,\, \bm{x}_n, \mu) \cr
 & = \int \dd^d \bm{y}
  \langle O(\bm{x}_1) \cdots O(\bm{x}_n) O(\bm{y}) \rangle_{\mu,\, 0}
 \prod^n_{i=1} \theta \left( |\bm{x}_i - \bm{y}| - 1/\mu \right) \,,
\end{align}
beginning at $\mu=\mu_0$ as in Sec.~\ref{Sec:RG}. Here, using the
Heaviside function, we explicitly described the condition
(\ref{Cond:cutoff}). Then, changing the cutoff from $\mu_0$ to $\mu$, we
obtain the change of $I_1^{(n)}$ as
\begin{align}
 & \Delta I_1^{(n)} (\bm{x}_1,\, \cdots,\, \bm{x}_n,\,\mu) \cr
 & = -  \int \dd^d \bm{y}
  \langle O(\bm{x}_1) \cdots O(\bm{x}_n) O(\bm{y}) \rangle_{\mu,\, 0}
 \sum^n_{i=1} \delta \left( |\bm{x}_i - \bm{y}| - 1/\mu \right) \Delta\!
 \left( 1 \over \mu \right)\,,
\end{align}
where we replaced the remaining Heaviside functions which we didn't take
differentiation with $1$. Taking the infinitesimal limit, we obtain
\begin{align}
  &\mu \frac{\dd}{\dd \mu} I_1^{(n)} (\bm{x}_1,\, \cdots,\,
 \bm{x}_n,\,\mu) \cr
  &\,\, =  \mu^{-1} \int \dd^d \bm{y}
  \langle O(\bm{x}_1) \cdots O(\bm{x}_n) O(\bm{y}) \rangle_{\mu,\, 0}
 \sum^n_{i=1} \delta \left( |\bm{x}_i - \bm{y}| - 1/\mu \right)\,. \label{tempI1n}
\end{align}
Using the OPE, given in Eq.~(\ref{OPE}), we rewrite the right-hand side as
\begin{align}
 & \langle O(\bm{x}_1) \cdots O(\bm{x}_n) O(\bm{y}) \rangle_{\mu,\, 0}\,
 \delta \left( |\bm{x}_1 - \bm{y}| - 1/\mu \right) \cr
 & \,= \frac{C}{c} \frac{1}{|\bm{x}_1 - \bm{y}|^{\Delta_0}} \langle O(\bm{x}_1) \cdots O(\bm{x}_n) \rangle_{\mu,\, 0}\,
 \delta \left( |\bm{x}_1 - \bm{y}| - 1/\mu \right) + \cdots\,, \label{tempOPE}
\end{align}
where the first term in the OPE (\ref{OPE}) should be eliminated by
introducing the counter term and the ellipsis denotes the non-singular
terms in the limit $|\bm{x}_1 - \bm{y}| \to 0$. Here, $\Delta_0$ denotes the scaling dimension at
$\mu=\mu_0$. Then, integrating about $\bm{y}$, we obtain
\begin{align}
  \mu \frac{\dd}{\dd \mu} I_1^{(n)} (\bm{x}_1,\, \cdots,\,
 \bm{x}_n,\,\mu) = n \tilde{C} \mu^{\lambda} I^{(n)}_0(\bm{x}_1,\,
 \cdots,\, \bm{x}_n) + \cdots\,, \label{Eq:In1}
\end{align}
where we introduced $\tilde{C} \equiv {\rm Vol}(S^{d-1}) C/c$,
$\lambda \equiv \lambda(\mu_0)$, and 
\begin{align}
 I_0^{(n)} (\bm{x}_1,\, \cdots \bm{x}_n) \equiv \langle O(\bm{x}_1) \cdots
 O(\bm{x}_n) \rangle_{0}\,.
\end{align}

Note that the non-singular terms in the OPE will be written in
the form $|\bm{x}_i - \bm{y}|^p$ with a non-negative real number $p$,
which will be replaced with $\mu^{-p}$ after integrating about
$\bm{y}$. Therefore, the contributions from the non-singular terms will
be suppressed by $(r\mu)^{-(\Delta_0 + p)}$ compared with the first term
in Eq.~(\ref{Eq:In1}), where $r$ is a scale associated with $I_1^{(n)}$,
and hence we can neglect them in the large scale limit. Noticing the fact that
$I^{(n)}_0$ is the correlator for the CFT, which does not 
vary in the change of $\mu$, we can solve Eq.~(\ref{Eq:In1}) as
\begin{align}
 & I_1^{(n)}(\bm{x}_1,\, \cdots,\, \bm{x}_n,\,\mu) = n \tilde{C}\,
 \frac{\mu^{\lambda} - (f r)^{-\lambda}}{\lambda}\, I_0^{(n)}
 (\bm{x}_1,\, \cdots,\, \bm{x}_n) \,,   \label{Exp:I1n}
\end{align}
where $f$ is a dimensionless constant. Since the two-point function for the ``CFT'' with the cutoff $\mu$
should agree with the one for the exact CFT, when
the distance $x_{12} \equiv |\bm{x}_1 - \bm{x}_2|$ is much bigger than $1/\mu$, the
dimensionful parameter $r$ in $I_1^{(2)}$ should be $x_{12}$, which is the only
dimensionful quantity included in the two-point function for the CFT. Then, we can determine
$I_1^{(2)}(\bm{x}_1,\,\bm{x}_2,\,\mu)$ up to the constant parameter $f$
as
\begin{align}
 & I_1^{(2)}(\bm{x}_1,\,\bm{x}_2,\,\mu) = 2 \tilde{C}\,
 \frac{\mu^{\lambda} - (f x_{12})^{-\lambda}}{\lambda}\, I_0^{(2)}
 (\bm{x}_1,\,  \bm{x}_2) \,.
\end{align}
By contrast, for $n \geq 3$, $I_1^{(n)}$ can depend on all the
distances $x_{ij} \equiv |\bm{x}_i - \bm{x}_j |$ with $i,j=1,\, \cdots,\, n$ and hence
we cannot determine the term $(fr)^{-\lambda}$ only from the dimensional
analysis. In the following, simply assuming that $(fr)^{-\lambda}$
should be expressed by a product of $x_{ij}$s, we use the formal
expression (\ref{Exp:I1n}).

Next, we compute $I^{(n)}_m(\bm{x}_1,\,\cdots,\,\bm{x}_n,\,\mu)$. 
Following a similar argument, we obtain 
\begin{align}
  &\mu \frac{\dd}{\dd \mu} I_m^{(n)} (\bm{x}_1,\, \cdots,\,
 \bm{x}_n,\,\mu) \cr
  & =  \mu^{-1} \int \dd^d \bm{y}_1 \cdots \int \dd^d \bm{y}_m 
  \langle O(\bm{x}_1) \cdots O(\bm{x}_n) O(\bm{y}_1) \cdots O(\bm{y}_m)
 \rangle_{\mu,\, 0} \cr
  & \qquad \qquad \times \left[
 \sum^n_{i=1} \sum_{j=1}^m \delta \left( |\bm{x}_i - \bm{y}_{j}| -
 1/\mu \right) +  \sum_{j,\, j'}   \delta \left( |\bm{y}_j - \bm{y}_{j'}| -
 1/\mu \right)  \right] \,, \label{Eq:dImn0}
\end{align}
where the second term in the last line represents all the combinations
about $j$ and $j'$ with $j, j'= 1,\, \cdots,\, m$. Then, using the OPE
(\ref{OPE}) and integrating about one of $\bm{y}_j$ or $\bm{y}_{j'}$ which appears in each
accompanied delta function, we obtain
\begin{align}
 & \mu \frac{\dd}{\dd \mu} I_m^{(n)}(\bm{x}_1,\, \cdots,\,
 \bm{x}_n,\,\mu) = B^{(n)}_m\, \tilde{C} \mu^{\lambda} \,
 I_{m-1}^{(n)} (\bm{x}_1,\, \cdots,\, \bm{x}_n,\, \mu)  \,, \label{Eq:dImn}
\end{align}
where we again neglected the non-singular contributions in the
OPE and $B^{(n)}_m$ denotes the number of terms with the delta functions
in the last line of Eq.~(\ref{Eq:dImn0}), given by
\begin{align}
 & B^{(n)}_m \equiv \frac{m}{2} (2n + m-1)\,.
\end{align}
Introducing $y(\mu)$ which satisfies
\begin{align}
 & \frac{\dd y(\mu)}{\dd \mu} = \tilde{C} \mu^{\lambda -1}\,, \label{Eq:chi}
\end{align}
we rewrite Eq.~(\ref{Eq:dImn}) as
\begin{align}
 & \frac{\dd}{\dd y} I_m^{(n)}(\bm{x}_1,\, \cdots,\,
 \bm{x}_n,\,y) =   B^{(n)}_m\, I_{m-1}^{(n)} (\bm{x}_1,\, \cdots,\,
 \bm{x}_n,\, y) \,. \label{Eq:dImn2}
\end{align}
Operating $\dd^{m-1}/\dd y^{m-1}$, we obtain
\begin{align}
 & \frac{\dd^m}{\dd y^m} I_m^{(n)}(\bm{x}_1,\, \cdots,\,
 \bm{x}_n,\,y) = \prod_{m'=1}^m  B^{(n)}_{m'}\, I_{0}^{(n)} (\bm{x}_1,\, \cdots,\,
 \bm{x}_n) \,.
\end{align}
Note that, since $I^{(n)}_0$ is the $n$-point functions for the CFT, we
can solve this equation as
\begin{align}
 & I_m^{(n)}(\bm{x}_1,\, \cdots,\, \bm{x}_n,\,y)
  = \frac{1}{m!} \prod_{m'=1}^m  B^{(n)}_{m'}\, I_{0}^{(n)} (\bm{x}_1,\, \cdots,\,
 \bm{x}_n) y^m + \sum_{m'=0}^{m-1} d_{m'} y^{m'}
\,, \label{Sol:dImn}
\end{align}
where $d_{m'}$ is constant in $y$, but can depend on $\bm{x}_i$ with
$i=1,\,\cdots,\, n$. Inserting this solution into
Eq.~(\ref{Eq:dImn2}), we obtain
\begin{align}
 & \sum_{m'=0}^{m-2} (m'+1) d_{m'+1} y^{m'} = B^{(n)}_m
 \sum_{m'=0}^{m-2} d_{m'} y^{m'} 
\,.
\end{align}
Comparing the coefficients of $\chi^{m'}$, we obtain
\begin{align}
 & d_{m'} = B^{(n)}_m \frac{d_{m'-1}}{m'} = \left\{ B^{(n)}_m
 \right\}^{m'} \frac{d_0}{m'!} 
\,.
\end{align} 
Inserting this expression into Eq.~(\ref{Sol:dImn}), we can solve  
\begin{align}
 & I_m^{(n)}(\bm{x}_1,\, \cdots,\, \bm{x}_n,\,y) \cr
 & = \frac{(2n+m-1)!}{(2n-1)!} I_{0}^{(n)} (\bm{x}_1,\, \cdots,\,
 \bm{x}_n) \left( y \over 2 \right)^m + d_0 \sum_{m'=0}^{m-1} \frac{1}{m'!}  \left\{
 B^{(n)}_m y \right\}^{m'} 
\,,  \label{Sol:dImn2}
\end{align}
where we used
\begin{align}
 &  \prod_{m'=1}^m  B^{(n)}_{m'}\, = \frac{m!}{2^m} \frac{(2n+m-1)!}{(2n-1)!}\,.
\end{align}
Solving Eq.~(\ref{Eq:chi}), we obtain
\begin{align}
 & y(\mu) = \tilde{C} \frac{\mu^{\lambda} - (fr)^{-\lambda}}{\lambda}\,,
\end{align}
where again we introduced $(fr)^{-\lambda}$ as an integration constant.

In the following, we set the integration constant $d_0$ to $d_0=0$, which
leads to $d_n=0$ with $n\geq 1$. Then, inserting Eq.~(\ref{Sol:dImn2})
into the $n$-point function of $O$, we obtain
\begin{align}
 & \langle O(\bm{x}_1) \cdots O(\bm{x}_n) \rangle_{\mu} \cr
 & \equiv \frac{1}{(2n-1)!}
 I_{0}^{(n)} (\bm{x}_1,\, \cdots,\, \bm{x}_n) 
  \sum_{m=0}^\infty  \frac{(2n+m-1)!}{m!}  \left( - \frac{u y}{2} \right)^m\,.
\end{align}
Using the negative binomial formula
\begin{align}
 & \sum_{m=0}^\infty \frac{(l+m)!}{l!\, m!} X^m = (1 -X)^{-(l+1)}\,, \label{Eq:binomial}
\end{align}
which is valid for a real number $l$ and $|X|<1$, we obtain
\begin{align}
 \langle O(\bm{x}_1) \cdots O(\bm{x}_n) \rangle_{\mu} =
 I_{0}^{(n)} (\bm{x}_1,\, \cdots,\, \bm{x}_n) \left[ 1 + \frac{u(\mu)y(\mu)}{2} \right]^{-2n} \,.
\end{align}

Using the solution of $g(\mu)$, given in Eq.~(\ref{Sol:gs}), we can
rewrite the $\mu$ dependent term in the square brackets as
\begin{align}
 &  \frac{u(\mu) y(\mu)}{2} =  \frac{g(\mu)
 \mu^{-\lambda} y(\mu)}{2} = \frac{ (f p r)^{-\lambda} - \Bigl( \frac{\mu}{p}
 \Bigr)^{\lambda}}{1 + \Bigl( \frac{\mu}{p}
 \Bigr)^{\lambda}}\,,
\end{align}
and hence we obtain
\begin{align}
 & \langle O(\bm{x}_1) \cdots O(\bm{x}_n) \rangle_{\mu} \cr
 &= I_{0}^{(n)} (\bm{x}_1,\, \cdots,\, \bm{x}_n) \left[ 1 +  \Bigl(
 \frac{\mu}{p}\Bigr)^{\lambda}  \right]^{2n} \left[ 1 +  (fpr)^{-\lambda}  \right]^{-2n}\,. 
\end{align}
Since the boundary operator ${\cal O}(\bm{x})$ is related to
$O(\bm{x})$ as ${\cal O}(\bm{x})= \mu^{-\lambda} O(\bm{x})$, the
correlator of ${\cal O}(\bm{x})$ is given as 
\begin{align}
 & \langle {\cal O}(\bm{x}_1) \cdots {\cal O}(\bm{x}_n) \rangle_{\mu} \cr
 &= I_{0}^{(n)} (\bm{x}_1,\, \cdots,\, \bm{x}_n) \mu^{- n \lambda} \left[ 1 +  \Bigl(
 \frac{\mu}{p}\Bigr)^{\lambda}  \right]^{2n} \left[ 1 +  (fpr)^{-\lambda}  \right]^{-2n}\,. 
\end{align}

\section{Vertex function from Ward-Takahashi identity}  \label{Sec:WTidentity}
In Sec.~\ref{SSec:WnCP}, we derived the expression for the vertex function
$W^{(n)} (\bm{x}_1,\, \cdots,\, \bm{x}_n)$ written by the $n$-point functions of
${\cal O}(\bm{x})$ in the flat space, using the relation
(\ref{Rel:zetadphi}). In this section, we address this relation in an
alternative way.

\subsection{Ward-Takahashi identity}  \label{SSec:WTidentity}
To derive the expression of the vertex function, following Ref.~\cite{JYsingle}, we derive the Ward-Takahashi
identity associated with the Weyl scaling which changes the spatial metric as 
\begin{align}
 & h_{ij}(\bm{x}) \to e^{-2\alpha(\sbm{x})} h_{ij}(\bm{x})\,. 
\label{Exp:Weyl}
\end{align}
In the gauge with Eq.~(\ref{Exp:gauge}), the Weyl transformation of the 
metric renders the curvature perturbation shifted as
\begin{align}
 & \zeta(\bm{x}) \to \zeta_\alpha(\bm{x}) \equiv \zeta(\bm{x}) -\alpha(\bm{x})\,.
\end{align}

The Ward-Takahashi identity stems from the following identity
\begin{align}
 & \int D\chi e^{- S_{\rm QFT}[
 \zeta_\alpha,\,\chi]} =\int D\chi_\alpha e^{-  S_{\rm
 QFT}[\zeta_\alpha,\,  \chi_\alpha]}\,,   \label{Exp:Z0}
\end{align}
which states that the generating functional 
$Z_{\rm QFT}[\zeta]$ should be independent of a choice of
integration variable. Here, we express the field $\chi$ after the
dilatation scaling as $\chi_\alpha$. Assuming that the integration
measure $D\chi$ is invariant under the Weyl scaling, we
consider the case with ${\cal J}[\alpha]=1$, where ${\cal J}[\alpha]$ is
the Jacobian, $D\chi_\alpha = {\cal J}[\alpha] D \chi$. (This case
corresponds to the case without the trace anomaly.) We will show that
once we provide
\begin{align}
 & \delta S_\alpha [\zeta,\, \chi] \equiv S_{\rm QFT}[\zeta_\alpha,\,  \chi_\alpha] - S_{\rm QFT} [\zeta,\,
 \chi] \label{Def:deltaS}\,,
\end{align}
which describes the change of the action $S_{\rm QFT}$ due to the Weyl
scaling, we can determine $W_{\rm QFT}[\zeta]$. Operating 
$\delta^n/\delta \alpha(\bm{x}_1) \cdots \delta \alpha(\bm{x}_n)$ on the
both sides of Eq.~(\ref{Exp:Z0}), dropping the contributions from
disconnected diagrams, and setting all 
$\alpha(\bm{x})$ and $\zeta(\bm{x})$ to $0$, we obtain 
\begin{align}
 &  \frac{\delta^n \ln Z_{\rm
 QFT}[\zeta] }{\delta \zeta(\bm{x}_1) \cdots \delta
 \zeta (\bm{x}_n)} \bigg|_{\zeta=0} 
\cr &\quad \qquad  = (-1)^n \left\langle \frac{\delta^n}{\delta \alpha(\bm{x}_1) \cdots \delta
 \alpha(\bm{x}_n)}  e^{-   \delta S_\alpha  [\zeta,\,  \chi]}
 \bigg|_{\alpha=0} \right\rangle_{\mu}\,,  \label{WTodd}
\end{align}
where we introduced
\begin{align}
 & \langle X[\chi] \rangle_\mu 
 \equiv {\int D \chi X[\chi] e^{  - S_{\rm
 QFT}[\zeta=0,\,\chi]} \over \int D \chi
 e^{  - S_{\rm QFT}[\zeta=0,\, \chi]} }\,. 
\end{align}
In deriving Eq.~(\ref{WTodd}), we noted that since the left hand side of
Eq.~(\ref{Exp:Z0}) includes $\alpha(\bm{x})$ only 
in the combination of $\zeta_{\alpha}(\bm{x}) =\zeta(\bm{x}) - \alpha(\bm{x})$, a derivative with respect to
$\alpha(\bm{x})$ can be replaced with a derivative with respect to $\zeta(\bm{x})$ as
$$
\frac{\delta}{\delta \alpha(\bm{x})} (\cdots)  = - \frac{\delta}{ \delta
\zeta(\bm{x})} (\cdots)\,.
$$

Inserting the Ward-Takahashi identity (\ref{WTodd}) into Eq.~(\ref{Def:Wn}), we can express the vertex function 
$W^{(n)} (\bm{x}_1,\, \cdots,\, \bm{x}_n)$ in terms of the
$m$-point functions with $m\leq n$ for ${\cal O}(\bm{x})$. For $n=1$
and $n=2$, we obtain
\begin{align}
 W^{(1)}(\bm{x}) &
 = - 2 {\rm Re} \left[   \left\langle \delta^1 S_{\alpha}(\bm{x}) \right\rangle_\mu \right]\,, \label{Exp:Z01}
\end{align}
and 
\begin{align} 
  &W^{(2)} (\bm{x}_1,\, \bm{x}_2)  =  - 2 {\rm Re} \left[
 \left\langle \delta^1 S_\alpha(\bm{x}_1) \, \delta^1 S_\alpha(\bm{x}_2)
 \right\rangle_\mu - 
 \left\langle  \delta^2 S_\alpha(\bm{x}_1,\, \bm{x}_2)  \right\rangle_\mu \right]\,,  \label{Exp:Z02} 
\end{align}
where we introduced the abbreviated notation
\begin{align}
 & \delta^n S_{\alpha} (\bm{x}_1,\, \cdots,\, \bm{x}_n) \equiv
 \frac{\delta^n S_\alpha[\zeta,\,  \chi ]}{\delta
 \alpha(\bm{x}_1) \cdots \delta \alpha(\bm{x}_n)} \bigg|_{\zeta=\alpha=0} \,.  \label{Def:dSa}
\end{align}
An extension to a general $n$ proceeds in a straightforward manner and gives 
\begin{align}
W^{(n)} (\bm{x}_1,\, \cdots,\, \bm{x}_n)
&= -2 {\rm Re}
 \Bigl[ \left\langle \delta^1 S_\alpha(\bm{x}_1) \cdots \delta^1 S_\alpha(\bm{x}_n) \right\rangle_\mu \cr
& \qquad   \qquad \, - \bigl\{ \! \left\langle \delta^2 S_\alpha(\bm{x}_1,\, \bm{x}_2)
 \delta^1 S_\alpha(\bm{x}_3) \cdots \delta^1 S_\alpha (\bm{x}_n)
 \right\rangle_\mu \hspace{-5pt} + \left( \cp \right)  \bigr\} \cr
 & \qquad   \qquad \, + \cdots + (-1)^{n-1} \left\langle \delta^n
 S_\alpha(\bm{x}_1,\, \cdots,\, \bm{x}_n ) \right\rangle_\mu \, \Bigr] \,. \label{Exp:Z0n}
\end{align}

\subsection{Boundary theory}  \label{SSec:ambiguity}
It will be useful to point out that the generating functional 
$W_{\rm QFT}[\zeta]$ can be
described also by using the energy momentum tensor in the curved
background. Actually, we can obtain the derivative of 
$\ln Z_{\rm QFT}[\zeta]=-W_{\rm QFT}[\zeta]$ with respect to $\zeta$, as 
\begin{align}
 \frac{\delta \ln Z_{\rm QFT}[\zeta]}{\delta \zeta(\bm{x})}
 \bigg|_{\zeta=0} &= - \langle T(\bm{x}) \rangle_{\mu}\,, \\
 \frac{\delta^2 \ln Z_{\rm QFT}[\zeta]}{\delta \zeta(\bm{x}_1)
 \delta \zeta(\bm{x}_2)} \bigg|_{\zeta=0} &=  \langle T(\bm{x}_1)
 T(\bm{x}_2) \rangle_{\mu} \cr
 & \quad - d\, \delta(\bm{x}_1 - \bm{x}_2)
 \langle T(\bm{x}_1) \rangle_{\mu} -  \left\langle \frac{\delta
 T(\bm{x}_1)}{\delta \zeta(\bm{x}_2)} \bigg|_{\zeta=0} \right\rangle_{\mu}\,,
\end{align}
and so on and using these expressions, we can express the perturbative
 expansion of $W_{\rm QFT}[\zeta]$ regarding $\zeta$.  Here,
 $T(\bm{x})$ denotes the trace part of the energy momentum tensor
 $T(\bm{x}) \equiv h^{ij}(\bm{x}) T_{ij}(\bm{x})$. Namely, comparing the expression of $\ln Z_{\rm QFT}$ described by the correlators of the energy momentum tensor to Eq.~(\ref{WTodd}), we find
\begin{align}
 & T(\bm{x}) = - \delta ^1 S_\alpha(\bm{x})  \,. \label{Exp:T}
\end{align}

In general, to obtain the
generating functional $W_{\rm QFT}[\zeta]$, which gives the wave
function $\psi[\zeta]$, we need to specify the boundary QFT in the
presence of the external field $\zeta$ or more explicitly we need to specify the energy momentum
tensor of the corresponding QFT. Meanwhile, the Ward-Takahashi identity (\ref{WTodd}) shows that when the change of the action 
$\delta S_\alpha  [\zeta,\,  \chi]$ under the Weyl scaling is specified, without consulting the detail of the boundary theory, we
can derive the generating functional $W_{\rm QFT}[\zeta]$
described by the $n$-point functions of ${\cal O}(\bm{x})$ in the
flat space. That is to say, the ambiguity in extending the QFT to
include the non-vanishing external field $\zeta$ can be attributed to the ambiguity in  
$\delta S_\alpha  [\zeta,\, \chi]$.

\subsection{Vertex function in a local boundary theory}  \label{SSec:simple}
When we consider a local theory as the boundary QFT, we can
naturally assume that in the limit $\zeta$ goes to $0$,  
$\delta^n S_{\alpha} (\bm{x}_1,\, \cdots,\, \bm{x}_n)$ are expressed 
by the boundary operator ${\cal O}$ as
\begin{align}
 & \delta^1 S_{\alpha} (\bm{x}) =   \beta_1(\mu) {\cal O}(\bm{x})
\end{align}
and for $n\geq 2$ as
\begin{align}
 & \delta^n S_{\alpha} (\bm{x}_1,\, \cdots,\, \bm{x}_n) =  (-1)^{n-1}  \beta_n(\mu)\,
 \delta(\bm{x}_1 - \bm{x}_2) \cdots \delta(\bm{x}_{n-1} - \bm{x}_n)\,
 {\cal O}(\bm{x}_1)   \label{Exp:dSa}
\end{align}
with the $\mu$ dependent coefficient $\beta_n$. Here, we noted that since the boundary theory is local, taking the derivative of the
boundary action $S_{\rm QFT}$ with respect to both $\alpha(\bm{x}_1)$ and
$\alpha(\bm{x}_2)$ will give the delta function 
$\delta(\bm{x}_1 - \bm{x}_2)$. In addition, after we set $\zeta$ to
$0$, the $\bm{x}$ dependent variable is only $\chi$ and hence the
$\bm{x}$ dependence should be described by the boundary operators ${\cal O}$
which is the composite operators of $\chi$. Then, inserting
Eq.~(\ref{Exp:dSa}) into Eq.~(\ref{Exp:Z0n}), we obtain
\begin{align}
 &W^{(n)}(\bm{x}_1,\, \cdots,\, \bm{x}_n) \cr
&= -2 {\rm Re}
 \Bigl[ \beta_1^n  \langle {\cal O}(\bm{x}_1) \cdots   
 {\cal O}(\bm{x}_n) \rangle_\mu \cr & \qquad \quad \qquad
  +  \beta_2  \beta_1^{n-2} \left\{ \delta (\bm{x}_1 - \bm{x}_2)  \langle {\cal O}(\bm{x}_2) \cdots  
 {\cal O}(\bm{x}_n) \rangle_\mu  + \left( \cp \right)  \right\} \cr
 & \qquad \quad  \qquad  + \cdots +  \delta
 (\bm{x}_1 - \bm{x}_2) \cdots \delta(\bm{x}_{n-1} - \bm{x}_n)
 \beta_n \langle {\cal O}(\bm{x}_1) \rangle_\mu  \Bigr] \,, \label{Exp:Z0gnap}
\end{align}
which is in the same form as Eq.~(\ref{Exp:Z0gn}). 
In Sec.~\ref{SSec:WnCP}, the parameters $\beta_n$ are determined by
using the relation between $\zeta$ and $\delta g_f$, (\ref{Rel:zetadphi}).

It will be instructive to observe how the function $\beta_n$ is given in a simple example. As such example, here, we
consider the boundary theory whose action is given by
\begin{align}
 & S_{\rm QFT}[\zeta,\, \chi] =  S_{\rm CFT}[\zeta,\, 
 \chi] -  \int \dd^d \bm{x} e^{d \zeta(\sbm{x})} g(\mu)  
 {\cal O}(\bm{x}) \,,
\end{align}
where the first term preserves the invariance under the Weyl
 transformation which changes $\chi(\bm{x})$ into
\begin{align}
 & \chi_\alpha (\bm{x}) = e^{\Delta_\chi \alpha(\sbm{x})} \chi(\bm{x})\,, 
\end{align}
with the scaling dimension $\Delta_\chi$
{\it i.e.,} the action $S_{\rm CFT}$ satisfies 
\begin{align}
 & S_{\rm CFT}[\zeta,\,\chi] = S_{\rm CFT}[\zeta - \alpha,\,
 e^{\Delta_\chi \alpha} \chi]\,.  \label{tempWeyl}
\end{align}
For simplicity, we consider the case where $\Delta_\chi$ remains
constant (at least at the energy scale $\mu$ we are concerned), and the
boundary operator ${\cal O}$ is a power of $\chi$ as 
${\cal O} \propto \chi^p$. Then, the change of the boundary
operator is given by 
\begin{align}
 & {\cal O}^{(\alpha)} (\bm{x}) = e^{\Delta \alpha(\sbm{x})}
 {\cal O} (\bm{x})\,, \label{Exp:Oaex}
\end{align}
with $\Delta \equiv p \Delta_\chi$ and the change of the boundary action is given by
\begin{align}
 & \delta S_\alpha [\zeta,\,  \chi] =  \int \dd^d \bm{x} e^{d \zeta(\sbm{x})}  \left[
   e^{(\Delta -d) \alpha(\sbm{x})} - 1 \right] g(\mu)  {\cal
 O}(\bm{x})\,. \label{Exp:dSasimple}
\end{align}
In this simple example, $\beta_n$, introduced in Eq.~(\ref{Exp:dSa}) is given by
\begin{align}
 & \beta_n(\mu) = (-1)^{n-1}(\Delta - d)^n g(\mu) \,.  \label{Exp:betasimple}
\end{align}
Inserting this expression into Eq.~(\ref{Exp:Z0gnap}), we can express the vertex function $W^{(n)}(\bm{x}_1,\, \cdots,\, \bm{x}_n)$ by the $m$-point
functions with $m \leq n$ of ${\cal O}$ at the flat spacetime. In this
case, without consulting the relation (\ref{Rel:zetadphi}), we can
derive the expression of the vertex function.


\end{document}